\documentclass[fleqn,usenatbib]{mnras}

\usepackage{newtxtext,newtxmath}
\usepackage{amsmath}
\usepackage[T1]{fontenc}

\DeclareRobustCommand{\VAN}[3]{#2}
\let\VANthebibliography\thebibliography
\def\thebibliography{\DeclareRobustCommand{\VAN}[3]{##3}\VANthebibliography}

\usepackage{graphicx}	
\usepackage{amsmath}	
\usepackage{booktabs}
\usepackage{orcidlink}
\usepackage{bm}

\title[Disentangling galaxy physics]{\texttt{pop-cosmos}: Disentangling galaxy properties from observables using data-driven approaches}

\author[B. Van den Bussche et al.]
{Benedict Van den Bussche,$^{\orcidlink{0009-0005-5575-1121}\,1}$\thanks{E-mail: bv322@cam.ac.uk}
Sinan Deger,$^{\orcidlink{0000-0003-1943-723X}\,1}$
Hiranya V.\ Peiris,$^{\orcidlink{0000-0002-2519-584X}\,1,2,3}$
Stephen Thorp,$^{\orcidlink{0009-0005-6323-0457}\,1}$
\newauthor
Daniel J.\ Mortlock,$^{\orcidlink{0000-0002-0041-3783}\,4,5}$
Boris Leistedt,$^{\orcidlink{0000-0002-3962-9274}\,4}$
Anik Halder,$^{\orcidlink{0000-0002-0352-9351}\,1}$
Madalina N.\ Tudorache,$^{\orcidlink{0000-0002-7288-6627}\,1}$
\newauthor
and Gurjeet Jagwani$^{\orcidlink{0009-0004-7935-2785}\,1,6}$
\\
$^{1}$ Institute of Astronomy and Kavli Institute for Cosmology, University of Cambridge, Madingley Road, Cambridge, CB3 0HA, UK\\
$^{2}${Cavendish Laboratory, Department of Physics, University of Cambridge, JJ Thomson Avenue, Cambridge, CB3 0HE, UK}\\
$^{3}$ The Oskar Klein Centre, Department of Physics, Stockholm University, AlbaNova University Centre, SE 106 91 Stockholm, Sweden\\
$^{4}$ Astrophysics Group, Imperial College London, Blackett Laboratory, Prince Consort Road, London, SW7 2AZ, UK\\
$^{5}$ Department of Mathematics, Imperial College London, London, SW7 2AZ, UK\\
$^{6}$ Research Computing Services, University of Cambridge, Roger Needham Building, 7 JJ Thomson Ave, Cambridge CB3 0RB, UK
}

\date{Accepted XXX. Received YYY; in original form ZZZ}

\pubyear{\the\year{}}

\begin{document}
\label{firstpage}
\pagerange{\pageref{firstpage}--\pageref{lastpage}}
\maketitle

\begin{abstract}
The physical processes that shape a galaxy's spectrum are strongly degenerate in observations, obscuring which processes act independently. Leveraging the \texttt{pop-cosmos} generative galaxy population model, we investigate how many independent degrees of freedom the rest-frame optical SED contains. We use a $\beta$-variational autoencoder (VAE) to compress a 16-parameter stellar population synthesis (SPS) description into a disentangled latent representation interpreted through mutual information (MI). We find that five independent dimensions suffice, corresponding to stellar mass, recent star formation, dust, and two degrees of freedom in the ionization state of the gas. Stellar metallicity and stellar age are not among these primary drivers; their spectral effects are distributed across the others rather than independently encoded. By tying each dimension to specific spectral features, this decomposition breaks the star-formation--dust--metallicity degeneracies that limit broadband photometry, and recovers the physical conditions of the gas in typical star-forming galaxies more cleanly than the line-ratio diagnostics in standard use.
\end{abstract}




\section{Introduction}

A galaxy's light encodes information about the physical processes that produced it, but these processes leave entangled signatures that are hard to separate using limited observations. Spectroscopic observations can be highly informative, but are time-intensive to obtain and challenging to model accurately. Conversely, photometric observations are abundant but lose critical spectral information by integrating over broad bandpasses. To interpret either, stellar population synthesis (SPS) models serve as the primary theoretical framework for capturing the composite building blocks of a galaxy \citep[see reviews by][]{Tinsley80,conroy13,iyer25}. 

SPS models construct the spectral energy distribution (SED) of a galaxy from the 'bottom up.' A simple stellar population (SSP) is built by convolving the initial mass function (IMF) with stellar isochrones and spectra across a range of effective temperatures, luminosities, and metallicities. These SSPs are then integrated over a star formation history (SFH) and chemical evolution model to obtain a composite stellar population (CSP), to which  additional emission and attenuation components are applied, including interstellar medium (ISM) ionization, dust, and active galactic nuclei (AGN). Fitting the resulting model to observational data maps the observed light back to the underlying physical processes, and has proven successful at characterizing galaxies \citep{leja17, cappellari17}. While early SPS models relied on only four or five parameters \citep{sawicki98, lower20}, modern SED fitting codes utilize higher-dimensional parameter spaces to capture the full diversity of galaxies \citep{pacifici23}. 

Despite their success, these parameter spaces have internal degeneracies \citep{chevallard16} that complicate the mapping between physical properties and observables. For instance, dust attenuation and stellar metallicity both redden the optical continuum, mimicking the SED of a quenched galaxy \citep{wise96,guo11}. Similarly, massive young stars outshine older populations, obscuring their spectral signatures and systematically skewing stellar mass estimates. Breaking such degeneracies observationally requires either expanded wavelength coverage -- such as incorporating the infrared \citep{wang24b} -- or relying on high-resolution spectroscopy to isolate specific emission and absorption features \citep{worthey94,kauffmann03,gallazzi05, kewley19}.

One widely used diagnostic for differentiating the mechanisms shaping galaxies uses emission-line ratios \citep[as reviewed by][]{kewley19}, linking the relative strengths of specific emission lines to the physical processes driving them. However, such diagnostics require moderate to high-resolution spectra, decouple the gas physics from the underlying stellar population, and provide only a partial view of the overall astrophysics \citep{baldwin81, kewley01, kewley02, perez09, kewley19}. Rather than relying on a few isolated lines, data-driven spectral models attempt to use the full spectrum simultaneously. For instance, principal component analysis (PCA) decomposes galaxy observables into orthogonal components \citep{connolly95, yip04, Beck16,sharbaf23}. While PCA is simple and interpretable, it is restricted to linear relationships between galaxy properties and their SEDs, limiting it to broad spectral classification. By contrast, unsupervised machine learning models, such as autoencoders (AEs), capture the non-linear information content of galaxies within a low-dimensional latent space \citep{portillo20,teimoorinia22, melchior23, liang23, scourfield23, nicolaou26}. While AEs achieve high reconstruction accuracy, their latent spaces typically lack clear physical interpretability. A complementary approach was demonstrated by \citet{iwasaki23}, who used a variational AE (VAE) to compress Sloan Digital Sky Survey (SDSS) spectra into four latent variables, and then applied PCA to align these axes with interpretable dimensions. A number of studies have enforced disentanglement conditions during training, but without modifying the latent space afterwards, to enable direct physical interpretations in different astrophysical settings such as stellar spectra \citep{sedaghat21}, the halo mass function \citep{guo24}, dark matter halo density profiles \citep{lucie22,lucie24a, lucie24c}, cosmic microwave background temperature power spectra \citep{piras25}, matter power spectra \citep{Piras24}, and quasar spectra \citep{guarneri25}, but to date have not been applied to galaxy SEDs. 

In this work, we enforce disentanglement on the latent space of a variational AE
 trained on galaxy SEDs to determine the effective number of degrees of freedom shaping the rest-frame optical. By leveraging the \texttt{pop-cosmos} generative galaxy population model \citep{alsing20,alsing23,leistedt23,alsing24,thorp24, thorp25a, thorp25b, deger25}, we combine the physical labels of SPS models with the low-dimensional representation of a VAE to isolate distinct physical processes shaping the SED. We investigate the spectral mechanisms that enable this disentanglement and contrast them with the limitations of broadband photometry. Both individually and in combination, these derived latent dimensions provide new diagnostics for the underlying astrophysics.

The work is structured as follows: Section~\ref{sec:mock_data} details the synthetic galaxy catalogue, encompassing SPS parameters, SEDs, and photometry, along with a description of the underlying \texttt{pop-cosmos} generative galaxy population model. Section~\ref{sec:Methods} outlines the architecture of the VAE and the information-theoretic metric mutual information (MI) employed for data interpretation. Section~\ref{sec:Results} presents our findings, identifying the fundamental SED dimensions found by the VAE, their spectral signatures, and how these break photometric degeneracies between physical properties. Section~\ref{sec:discussions} integrates these results, comparing a new galaxy evolution diagnostic with established ones. We conclude in Section~\ref{sec:Ccl}.

\section{Sample generation} \label{sec:mock_data}
We draw our sample from a mock galaxy catalogue generated with the \texttt{pop-cosmos} generative model \citep{alsing24, thorp25b, deger25}, spanning the rest-frame ultraviolet (UV) to mid-infrared (MIR). This model was calibrated against the broad, intermediate, and narrow-band photometry of 423,262 galaxies in the COSMOS2020 catalogue \citep{weaver22} with a \textit{Spitzer} IRAC \textit{Ch.\,1} $< 26$ selection. Its wavelengths span from the CFHT $u$-band at $3450$~\AA\ to IRAC \textit{Ch.\,2} at $50127$~\AA. The \texttt{pop-cosmos} model provides the full joint distribution of galaxy physical properties and redshifts for $0 < z < 6$. These physical properties are the parameters of an SPS model, which maps them to a galaxy's SED.

The SPS model is defined by 16 physical parameters (Table \ref{tab:properties}), encompassing stellar and gas-phase metallicity, dust attenuation (utilizing the \citealt{calzetti00} attenuation law, and \citealt{charlot00} birth-cloud model), and a non-parametric SFH. This SFH is parametrized by star formation rate (SFR) across seven temporal bins, with the youngest two bins spanning 0–30~Myr and 30–100~Myr. We also use several derived parameters for our analysis. Remaining stellar mass is calculated by multiplying the formed stellar mass by a retention fraction, which is rapidly computed via a neural network emulator based on the 16 base parameters. Additionally, the mass-weighted age and the SFR averaged over the most recent 100 Myr are derived from the SFH parameter ratios, $\Delta\log_{10}(\mathrm{SFR})$ \citep{leja19_sfh, thorp25b}.

\begin{table}
    \centering
    \caption{SPS parameters (\textit{top}) and derived quantities (\textit{bottom}).}
    \label{tab:properties}
    \begin{tabular}{ll}
\toprule
symbol$/$unit & definition \\
\midrule
         $\log_{10}(M^{\mathrm{form}}/\mathrm{M}_{\odot})$ & stellar mass formed \\
         $\log_{10}(Z/\mathrm{Z}_{\odot})$ & stellar metallicity \\
$\Delta\log_{10}(\mathrm{SFR})_{\{1:6\}}$ & SFR ratios between adjacent SFH bins \\
$\tau_2/\mathrm{mag}$ & diffuse dust optical depth at 5500~\AA \\
$n$ & power law index for diffuse dust law  \\
$\tau_1/\tau_2$ & birth cloud dust optical depth  \\
$\ln(f_{\mathrm{AGN}})$ & AGN bolometric luminosity fraction  \\
$\ln(\tau_{\mathrm{AGN}})$ & AGN torus optical depth  \\
$\log_{10}(Z_{\mathrm{gas}}/\mathrm{Z}_{\odot})$ & gas-phase metallicity \\
$\log_{10}(U_{\mathrm{gas}})$ & gas ionization  \\
$z$ & redshift  \\
\midrule
$t_{\mathrm{age}}/\mathrm{Gyr}$ & mass-weighted age  \\
$\log_{10}(M/\mathrm{M}_{\odot})$ & stellar mass remaining  \\
$\log_{10}(\mathrm{SFR}/\mathrm{M}_{\odot}\,\mathrm{yr}^{-1})$ & SFR \\
$\log_{10}(\mathrm{sSFR}/\mathrm{yr}^{-1})$ & SFR per unit stellar mass remaining  \\
\bottomrule
     \end{tabular}
\end{table}
For each galaxy in the \texttt{pop-cosmos} mock galaxy catalogue, COSMOS2020 bandpass fluxes are generated using the Flexible SPS \citep[FSPS;][]{conroy09, conroy10a, conroy10b} and \texttt{Prospector} frameworks \citep{leja17, leja19,johnson21}. The underlying stellar population models employ MESA isochrones and stellar tracks \citep[MIST;][]{paxton11, paxton13, paxton15, dotter16, choi16} combined with empirical stellar templates from the MILES library \citep{sanchez06, falcon11}. Nebular continuum and emission lines are included using pre-computed \texttt{CLOUDY} photoionization grids \citep{ferland13, byler17}. This generation of photometry and emission-line fluxes is accelerated using \texttt{Speculator}, an emulator-accelerated SPS model \citep{alsing20} trained on FSPS and \texttt{Prospector} outputs. Realistic noise models are applied to the fluxes, before applying an IRAC \textit{Ch.\,1} $< 26$ selection cut. 

Our analysis relies on \texttt{pop-cosmos} having learned a robust connection between SPS properties and the SEDs they produce. This model has been extensively validated: in addition to stringent tests against COSMOS2020 data, it reproduces well-studied galaxy evolution trends. These include the star-forming main sequence, the mass-metallicity relation, and the fundamental metallicity relation \citep{thorp25b}. The model recovers high-confidence spectroscopic redshift distributions \citep{thorp24} and yields stellar mass functions consistent with \citet{weaver22, weaver23} across $0\leq z\leq 3.5$ for galaxies below $10^{11} \mathrm{M}_\odot$. \citet{deger25} found that it produces a cosmic star formation rate density with a steeper low-$z$ slope and a peak at $z = 1.3$, in better agreement with recent radio \citep{leslie20} and cosmic infrared background \citep{chiang25} measurements than standard literature estimates \citep[e.g.][]{madau14}. \citet{halder26} also demonstrated the model's capacity to infer properties for wide-field surveys on which it was not calibrated, with low redshift outlier fractions indicating robust generalization to unseen data. 

Since each SED follows deterministically from its SPS parameters, validating their population distribution validates the resulting SED population. We generate noiseless, rest-frame SEDs (in units of $\mathrm{L}_\odot\,\text{Hz}^{-1}$) for the \texttt{pop-cosmos} mock catalogue, by processing their SPS parameters using FSPS \citep{conroy09, conroy10a, conroy10b}, accessed via \texttt{python-fsps} \citep{pythonfsps} and the \texttt{Prospector} framework \citep{leja17, leja19, johnson21}. We extract the individual nebular emission-line luminosities calculated from the \texttt{CLOUDY} photoionization grids using \texttt{python-fsps} as in \cite{byler17}. We do not apply the emission line corrections from \texttt{pop-cosmos}. For all subsequent analyses, we restrict the rest-frame SEDs to the optical window between $2425$--$7327$ \AA. We use the default wavelength grid from FSPS, which includes 4223 wavelength points over the relevant window with an average resolution of $1.16$ \AA. 

From the full \texttt{pop-cosmos} mock catalogue and our corresponding FSPS SEDs, we draw a representative subset of 90,000 galaxies (SPS parameters, photometry, and SEDs) above the \citet{thorp25b} stellar mass completeness limit, and use these to train our models (Section \ref{sec:VAE}). We use normalized SEDs as our inputs, defined as the rest-frame SEDs divided by their median luminosity between $5300$--$5850$ \AA. 

\section{Methods} \label{sec:Methods}
To isolate the physical processes in galaxy SEDs, we use a VAE (Section \ref{sec:VAE}) to compress the SED into a minimal set of disentangled latent variables. We map these to known physical properties and specific spectral features using MI (Section \ref{sec:MI}). Together, these methods reveal which physical mechanisms act as the fundamental degrees of freedom shaping the rest-frame optical, and how these break the observational degeneracies that limit photometry. 

\subsection{Model architecture} \label{sec:VAE}
We train a $\beta$-VAE (shortened to VAE in this work; \citealt{higgins17}) on 90,000 mock galaxy SEDs. It is an unsupervised representation learning model consisting of an encoder and a decoder both parametrized by neural networks \citep[see review by][]{kingma19}. These networks are connected through a low-dimensional latent space, which acts as an information bottleneck, forcing the model to learn a compressed probabilistic representation of the input data (in our case, galaxy SEDs). The only input to the model are the generated normalized SEDs and their normalization factor and not the SPS parameters from which they were produced.

The encoder network, $q_\phi(\mathbf{z}|\mathbf{x})$, maps the input data $\mathbf{x}$ (a normalized SED and its normalization factor) to the parameters -- specifically, the mean $\bm{\mu}$ and log variance $\log(\bm{\sigma}^2)$ -- of a probabilistic latent distribution, assumed to be a multivariate normal distribution $\mathcal{N}[\bm{\mu}, \text{diag}(\bm{\sigma}^2)]$. The encoder processes the input through seven 1D convolutional layers starting with a filter size of 16, and doubling in size at each layer until 1024 filters total. The flattened output of the final convolutional layer is connected to a linear layer with 512 neurons, which is in turn connected to the 5-dimensional latent space. The decoder network, $p_\theta(\mathbf{x}|\mathbf{z})$, mirrors the encoder's architecture with transpose convolutional layers \citep{zeiler10} replacing the convolutional layers. It takes a sampled latent vector $\mathbf{z}$ as input from the 5-dimensional linear layer and is connected to a 512 neuron linear layer. This is then connected to transpose convolutional layers halving in size starting from 1024 filters to 16. Every convolutional and transpose convolutional layer has a stride of two, padding of one, and a kernel size of three. We do not use dropout; it delayed convergence and overfitting was not an issue with this model and our sample. We also do not use batch normalization in our configuration. We use LeakyRELU activations \citep{maas13} for all our layers with the negative slope set to 0.15, except for the final transpose convolutional layer of the decoder where we do not use an activation. We detail the hyperparameter tuning in the Appendix \ref{sec:annealing}. The decoder finally outputs a reconstructed normalized SED and normalization factor, denoted as $\mathbf{x}'$. The parameters $\phi$ and $\theta$ are learned jointly by minimising a total loss function. 

The loss function of the VAE comprises two terms: a mean squared error (MSE) reconstruction loss and a Kullback--Leibler (KL) divergence term.  The KL divergence measures the difference between the learned latent distribution $q_\phi(\mathbf{z}|\mathbf{x})$ and a prior distribution $p(\mathbf{z})$. It is defined as 
\begin{equation}
    D_{\rm KL}[q_\phi(\mathbf{z}|\mathbf{x}) || p(\mathbf{z})] = \int q_\phi(\mathbf{z}|\mathbf{x}) \ln \left( \frac{q_\phi(\mathbf{z}|\mathbf{x})}{p(\mathbf{z})} \right) \,\mathrm{d}\mathbf{z}.
\end{equation}
The prior $p(\mathbf{z})$ is chosen to be a multivariate normal distribution with independent unit variances, $\mathcal{N}(\mathbf{0}, \mathbf{I})$. Without the KL divergence term, the neural network minimizes the loss by folding mixtures of the physical properties into every available latent dimension, which is not conducive to scientific interpretation. The KL term, weighted by $\beta$ as below, is therefore necessary to disentangle the dimensions, so that the latent space can be interpreted as independent physical degrees of freedom.

The total loss is then given by the sum of the expected MSE and the KL divergence, 
\begin{equation}
    \mathcal{L}(\phi, \theta; \mathbf{x}) = \mathbb{E}_{q_\phi(\mathbf{z}|\mathbf{x})}[\mathrm{MSE}(\mathbf{x},\mathbf{x}')] + \beta D_{\rm KL}[q_\phi(\mathbf{z}|\mathbf{x}) || p(\mathbf{z})],
\end{equation} 
where $\mathbf{x}' \sim p_\theta(\mathbf{x}|\mathbf{z})$ is the reconstructed SED and normalization factor, and the expectation is taken over the latent distribution. Because the network reconstructs both the $n$ spectral data points and the single normalization factor, the MSE is averaged over all $n+1$ features. To account for the dynamic range of the normalization factor and its importance relative to the 4223 spectral points, we apply a weighting term $\alpha$ to its error. The expanded empirical loss is therefore
\begin{equation}
\begin{split}
    \mathcal{L} =& \frac{1}{n+1}\left[\sum_{i=1}^n(x_i-x_i')^2+ \alpha(x_{\rm norm}-x'_{\rm norm})^2\right] \\
    &+ \beta D_{\rm KL}[q_\phi(\mathbf{z}|\mathbf{x}) || p(\mathbf{z})].
\end{split}
\end{equation}

The parameter $\alpha$ is set to 10,000 such that the two terms in the MSE parentheses are roughly of the same order, ensuring the model learns a representation that preserves the information necessary for accurate reconstruction.

\begin{figure}
	\includegraphics[width=\linewidth]{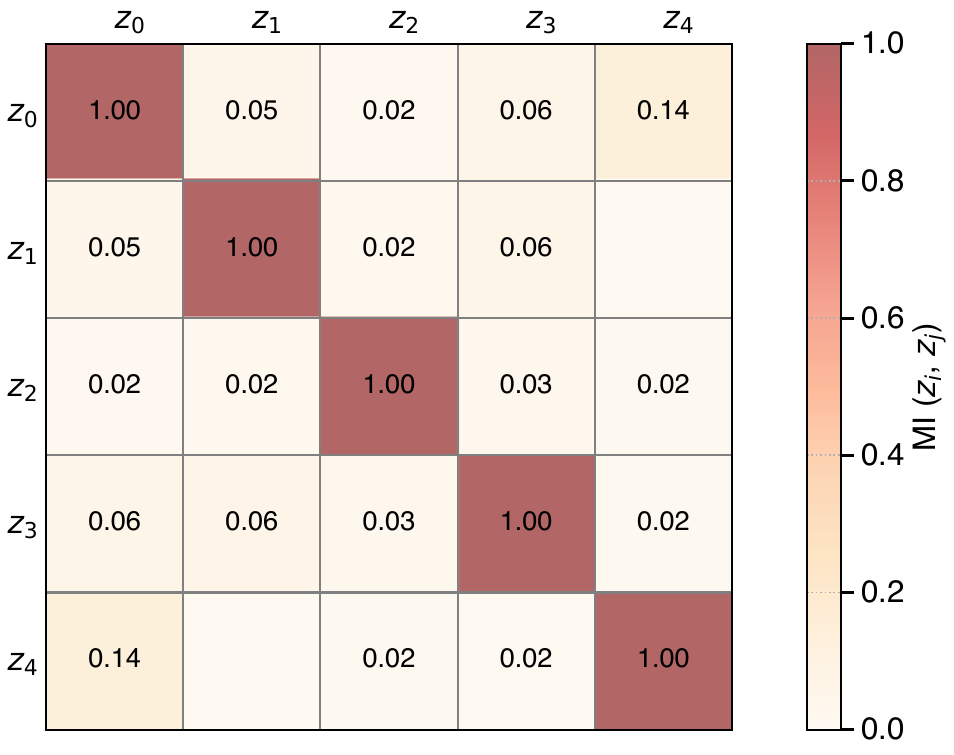}
    \caption{Disentanglement of the latents after training the VAE as measured by the MI between them. The MI is in natural units. The MI between identical latents are fixed to 1. MI values under 0.01 are not displayed. Uncertainty on the MI measurements is less than 0.001.}
    \label{fig:disentanglement}
\end{figure}

By satisfying this loss objective, a VAE can learn a representation of a galaxy SED. We control the disentanglement explicitly through the value of $\beta$ in the loss function. A higher $\beta$ encourages a more disentangled representation, in which latent dimensions capture independent factors of variation in the data. The degree of disentanglement can be quantitatively assessed by measuring the MI (see Section \ref{sec:MI}) between the latent variables themselves (Figure \ref{fig:disentanglement}); lower MI indicates a more disentangled and more interpretable representation. For meaningful scientific interpretation, the MI between latent variables should be lower than the MI between individual latent variables and the galaxy properties of interest. 

\begin{figure}
	\includegraphics[width=\linewidth]{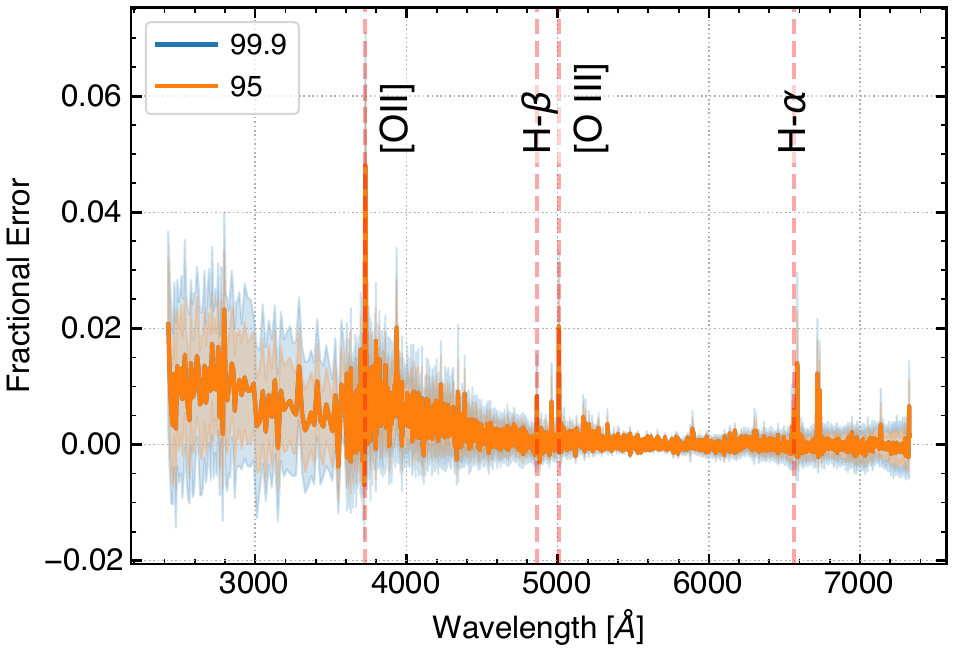}
    \caption{Normalized SED fractional reconstruction error after training the VAE. The error is evaluated on the validation set alone, and is calculated as the difference between the true and predicted SED divided by the true SED. Significant emission lines are highlighted in dashed red lines. The average overall error is $1.82\times 10^{-3}$ with the 95 percentile shaded in blue and the 99.9 percentile in orange.}
    \label{fig:reconstruction}
\end{figure}

We randomly partition the dataset of 90,000 mock galaxies (normalized SEDs and their normalization factors) into training (60,000 galaxies) and validation (30,000 galaxies) sets. We confirmed that the training set was large enough to saturate the model's accuracy. The reconstruction accuracy of the trained VAE is evaluated by quantifying the fractional difference between the input and output normalized SEDs from the validation set (see Figure~\ref{fig:reconstruction}), as well as between the input and output normalization factor. We achieve a two per cent fractional maximal error on the normalized SEDs, and less than 0.01 per cent on the normalization factor, ensuring that the model retains crucial physical information of the input SEDs. This error is maximal around significant emission lines ([O\,\textsc{ii}], [O\,\textsc{iii}], and H~$\upalpha$) and at shorter wavelengths where the spectral resolution is poorest. Above 4500 \AA\ however, the error falls closer to 0.01 per cent. 

To achieve both a disentangled latent space and high reconstruction accuracy, we employ a multistage $\beta$-annealing training process as in \citet{guo24}. This strategy involves gradually decreasing the value of $\beta$ during training, allowing the model to first learn a basic representation with a disentangled latent space, and then progressively focus on the reconstruction. The annealing strategy as well as the hyperparameter tuning is described in more detail in Appendix \ref{sec:annealing}.  

\subsection{Mutual information} \label{sec:MI}
MI measures the statistical dependence between two random variables, indicating the reduction in uncertainty about one variable given knowledge of the other \citep{shannon48a, shannon48b}.
For continuous random variables $X$ and $Y$ with a joint probability density function $p_{X,Y}(x,y)$ and marginal probability density functions $p_X(x)$ and $p_Y(y)$, the MI $I(X,Y)$ is given by
\begin{equation} \label{eq:MI}
    I(X,Y) = \int_{X\times Y}p_{X,Y}(x,y)\ln{\frac{p_{X,Y}(x,y)}{p_X(x)p_Y(y)}}\,{\rm d}x\,{\rm d}y .
\end{equation}
MI is measured in natural units (nat). A key property of MI is that $I(X,Y) = 0$ if and only if $X$ and $Y$ are statistically independent, as in this case $p_{X,Y}(x,y) = p_X(x)p_Y(y)$. 

We can also measure the conditional MI between $X$ and $Y$ given another random variable $Z$. If it increases compared to the MI between $X$ and $Y$, then $Z$ provides complementary information, whereas if it decreases then $Z$ provides redundant information. 

Traditional methods for estimating MI often involve binning the data or employing Bayesian inference \citep{hutter05}. However, these approaches are often sensitive to binning choices, prior assumptions, and hyperparameter tuning, particularly when applied to high-dimensional datasets with non-linear dependencies \citep{piras23}.

To estimate the MI, we utilize the publicly available GMM-MI algorithm and its associated Python package \citep{piras23}. GMM-MI employs Gaussian mixture models (GMMs) and bootstrapping to provide a reliable measure of the MI and its associated uncertainty. We initialize the model selection process using 3-fold cross-validation with 5 parameter initializations to determine the optimal number of GMM components to fit the distribution of the $x$ and $y$ samples in Equation \ref{eq:MI}. During the fitting process, we apply a covariance regularization of $10^{-5}$ and set the convergence threshold to $10^{-14}$ to ensure numerical stability. Finally, the MI distribution is estimated using $10^5$ Monte Carlo samples evaluated across 100 bootstrap iterations.
GMM-MI has already been used to investigate the latent space of deep learning models \citep{lucie24b, lucie24a, guarneri25}. The uncertainty on all our recorded MI is always at least an order of magnitude lower than the reported value, and often much lower. 

\section{Results} \label{sec:Results}
Based on both reconstruction accuracy and latent disentanglement, we find that five independent latent dimensions are required to describe the rest-frame optical SEDs of COSMOS2020-like galaxies drawn from the \texttt{pop-cosmos} galaxy population model. Adding a sixth dimension yields no improvement to the 99th percentile accuracy across the analysed wavelength range, whereas removing a dimension increases the fractional error by over two per cent. The model also disentangles this latent space, consistently achieving MI scores below 0.15 natural units (Figure \ref{fig:disentanglement}). While standard optical photometric diagnostics suffer from physical degeneracies \citep{conroy13}, the VAE isolates five degrees of freedom that break them using the full SED. We use MI to map these five latent variables to distinct physical drivers of the rest-frame optical: stellar mass ($z_4$), recent star formation ($z_0$), dust attenuation ($z_2$), and the soft ($z_1$) and hard ($z_3$) ionization potentials. The numbering of the latents is arbitrary and implies no ranking.

\subsection{Interpreting the latent diagnostics}
\begin{figure}
	\includegraphics[width=\linewidth]{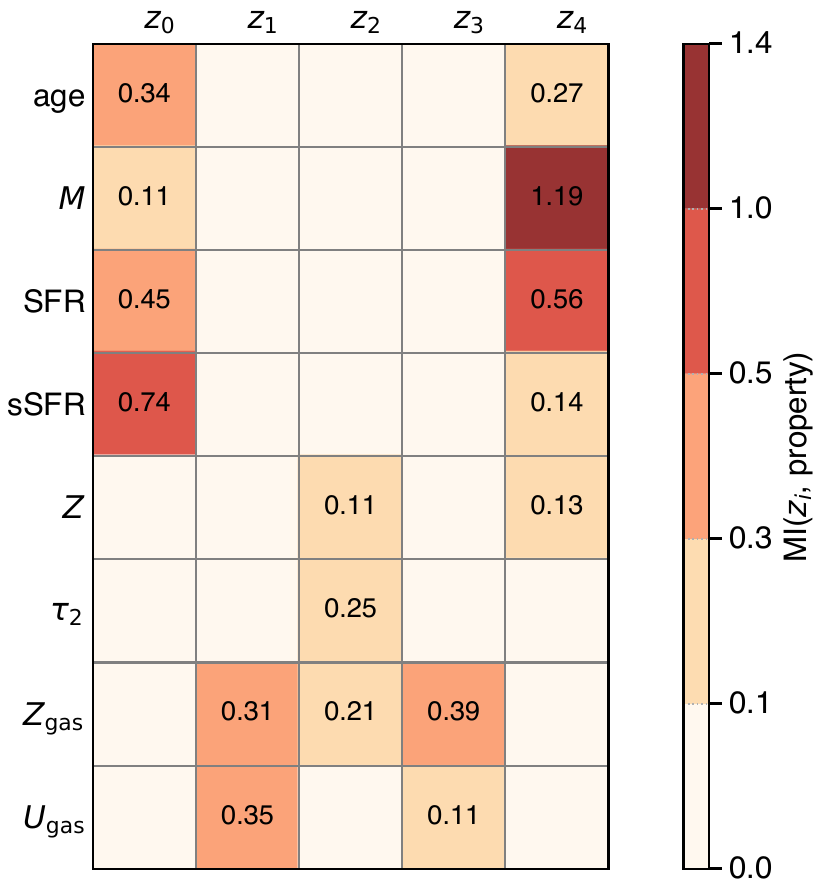}
    \caption{MI between the five latent variables of the VAE trained on rest-frame SEDs, and galaxy properties. MI values above 0.1 are overplotted on the heatmap. Only key galaxy SPS parameters isolated by the VAE are plotted on the heatmap. The uncertainties on MI values are less than 0.01. 
    }
    \label{fig:MI_lat_prop}
\end{figure}

\begin{figure}
    \centering
	\includegraphics[width=\linewidth]{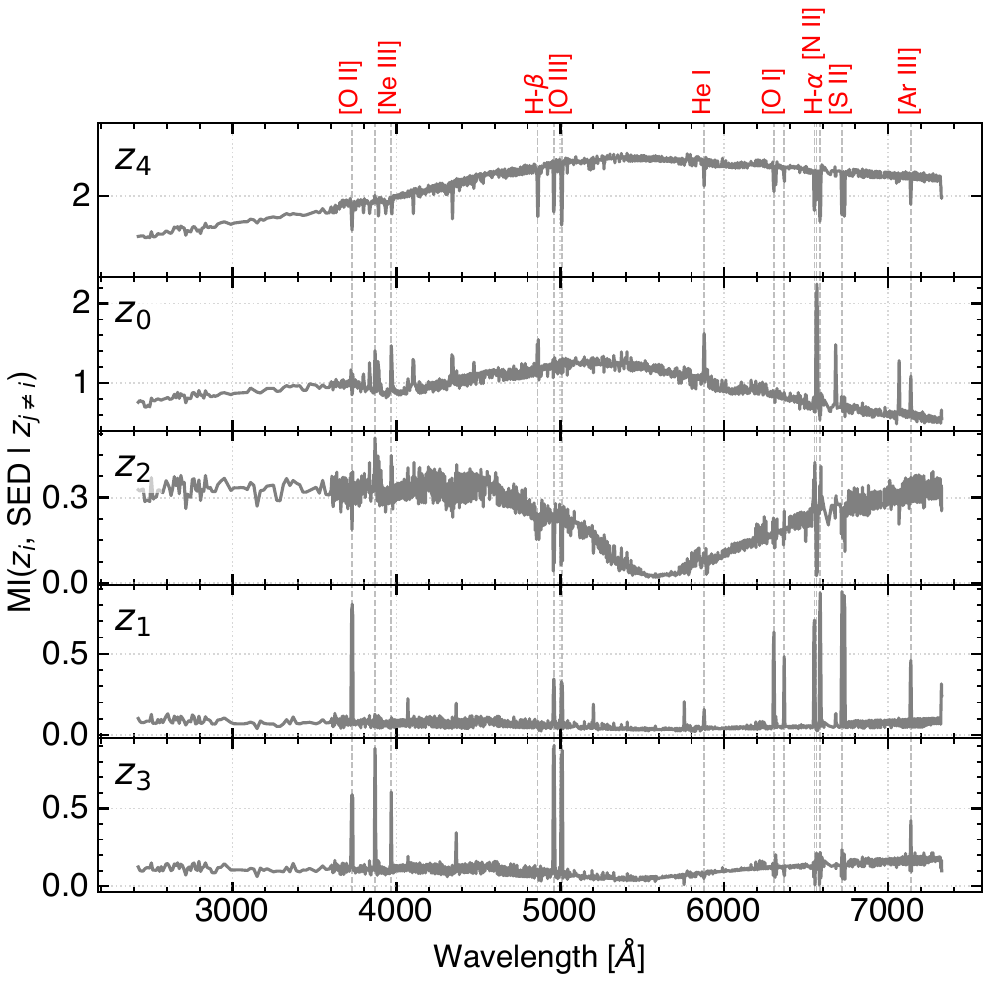}
    \caption{Conditional MI between individual latents and the rest-frame SED at each spectral indices, conditioned on all the other latents. Significant emission lines are overplotted in dotted gray lines with red labels above. The latents are ordered from \textit{top} to \textit{bottom} by decreasing average MI across the spectrum.} 
    \label{fig:MI_lat_SED_cond}
\end{figure}

\begin{figure*}
    \centering
	\includegraphics[width=\linewidth]{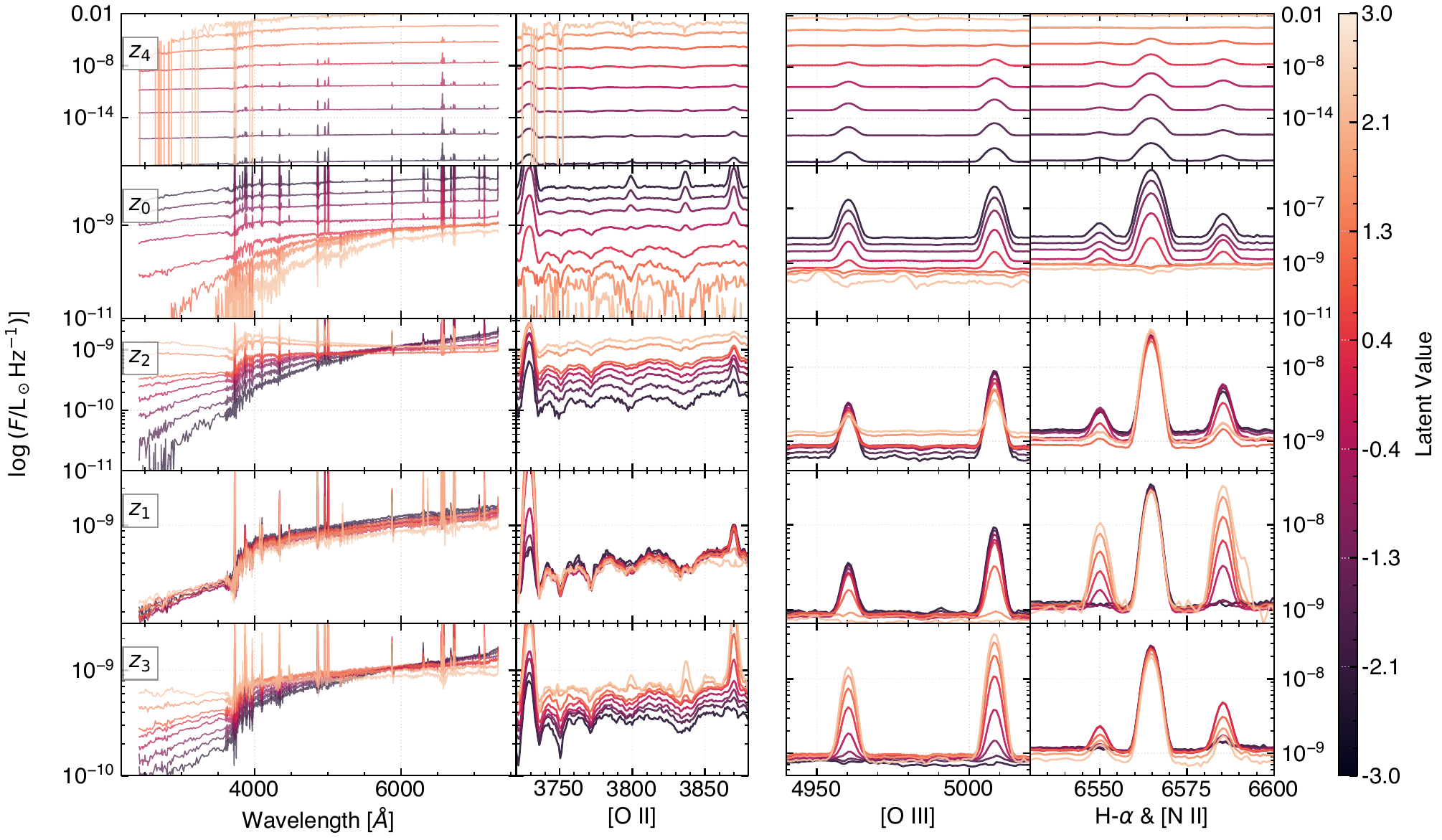}
    \caption{Variations in the decoded rest-frame SED when varying one latent eight times with all the others fixed. A different latent is traversed in each panel from top to bottom. The \textit{left} panel displays the full SED while the \textit{right} panel zooms into the [O\,\textsc{ii}], [O\,\textsc{iii}], H~$\upbeta$, and H~$\upalpha$ regions. The first two  and last two columns share a $y$-axis range. The different latents are ranked by their relative effect on the spectrum.} 
    \label{fig:lat_trav} 
\end{figure*}

Our interpretation rests on a set of MI-based diagnostics, which we introduce here before applying them to the individual latents in the following subsections. In Figure \ref{fig:MI_lat_prop} we measure the MI between latents and SPS parameters to determine which properties each latent is informative about. MI is measured in natural units (nats) throughout the paper; we will omit them for readability. Higher MI values indicate that a latent carries more information about a given property. We then identify the spectral features each latent controls by calculating the conditional MI between the latent and the SED as a function of wavelength (Figure \ref{fig:MI_lat_SED_cond}), conditioned on all the other latents. As discussed in Section \ref{sec:MI}, this conditioning isolates each latent's unique contribution, so in a disentangled space the resulting signals pick out distinct aspects of the SED. This is visually corroborated by the latent traversals in Figure \ref{fig:lat_trav}, which illustrate the isolated spectral effects of varying individual latents while holding all the others constant. Finally, we contrast the representations encoded in the VAE against the information accessible via observed-frame broad-band photometric colours across different redshift slices. We do this by measuring the MI between these colours and the SPS properties (Figure~\ref{fig:MI-photometry}). If a colour shares MI with two different SPS parameters, then those properties are degenerate within it; if it is informative about a single property, it is a good tracer of it. We also quantify the additional information about SPS properties that the VAE encodes from specific emission lines, which may otherwise be blended in standard photometry (Figure~\ref{fig:MI_emission}). 

\subsection{The stellar mass dimension} \label{sec:mass}
\begin{figure*}
    \centering
	\includegraphics[width=\linewidth]{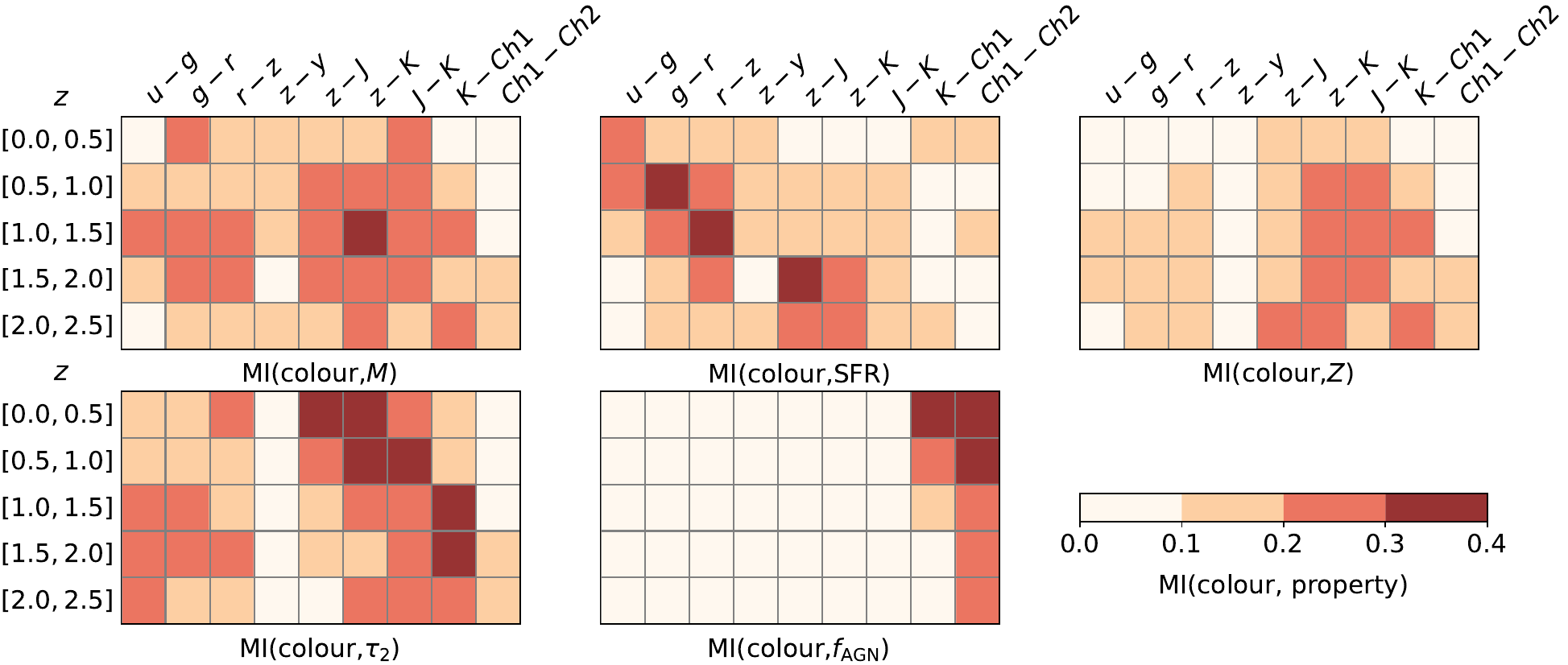}
    \caption{MI between different galaxy properties (\textit{top row:} stellar mass, sSFR, and stellar metallicity; \textit{bottom row:} diffuse dust optical depth $\tau_2$, and AGN bolometric luminosity fraction $f_{\rm AGN}$) and observed frame photometric colours for different redshift slices (vertical axis $z$). MI values are approximated to the closest decimal. The colourmap is discretized in bins of width 0.1. The uncertainty on the MI values are all less than 0.01.} 
    \label{fig:MI-photometry}
\end{figure*}

\begin{figure}
    \centering
    \includegraphics[width=\linewidth]{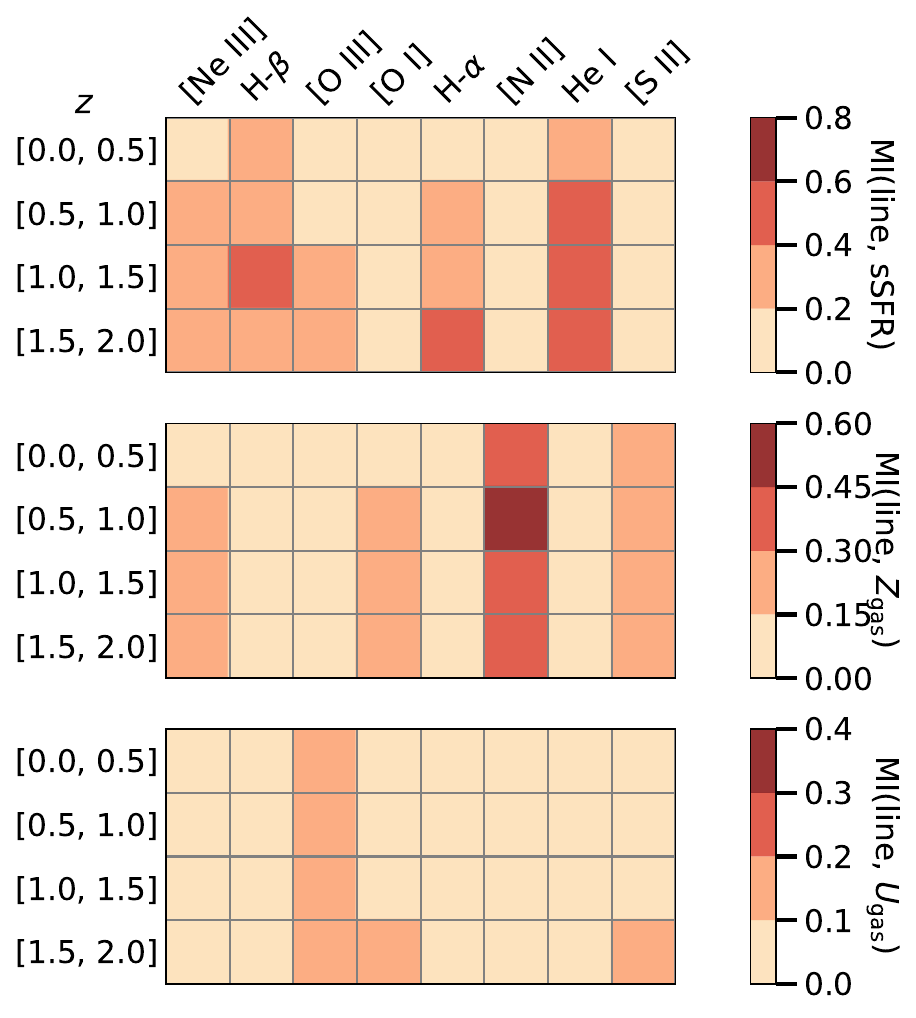}
    \caption{MI between the different galaxy properties (\textit{top} to \textit{bottom}: sSFR, gas-phase metallicity, and gas ionization) with different emission lines in rest-frame SEDs for galaxies in four different redshift slices.   
    } 
    \label{fig:MI_emission}
\end{figure}

The latent variable $z_4$ isolates the stellar mass: its MI with stellar mass exceeds that with any other property (Figure \ref{fig:MI_lat_prop}). Figure \ref{fig:MI_lat_SED_cond} shows that $z_4$ captures the SED normalization, as the conditional MI peaks at the wavelengths over which the normalization is defined ($5300$--$5850$ \AA). The normalization and the stellar mass are linked through the mass-to-light ratio ($M/L$) \citep{bell01, bruzual03}. While unobscured young stars dominate blue optical light, the optical--near-infrared (NIR) continuum reflects the aggregate stellar population \citep{bruzual03}. The MI for $z_4$ remains high at longer wavelengths beyond the normalization peak (Figure \ref{fig:MI_lat_SED_cond}), reflecting its sensitivity to this underlying population. The same long-wavelength continuum underlies broadband mass estimates, which rely on NIR colours to capture the continuum normalization (Figure \ref{fig:MI-photometry}).

Both $z_4$ and these colours also carry some information about stellar metallicity (Figure \ref{fig:MI_lat_prop}), which induces line blanketing that mimics changes in the $M/L$ ratio \citep{worthey94,courteau14}, creating a degeneracy with stellar mass. Figure \ref{fig:MI-photometry} shows that photometry can partially break this mass-metallicity degeneracy by incorporating the rest-frame UV continuum, visible in bluer broadband colours at higher redshifts. A steep UV continuum is linked to a galaxy population with low stellar mass and metallicity \citep{meurer99, salim18}, whose fractional abundance reflects the total stellar mass distribution. The VAE does not see the UV. Stellar metallicity is therefore not a key degree of freedom of the rest-frame optical; its effects appear jointly with those of dust and total stellar mass. 

\subsection{The young stellar population dimension} \label{sec:SFR}
Figure \ref{fig:MI_lat_prop} shows that recent star formation is encoded in $z_0$. It is most informative about the 4000 \AA\ break region -- $D_{n}$(4000) \citep{balogh99} -- and Balmer series (H~$\upalpha$, H~$\upbeta$, H~$\updelta$) as the conditional MI is substantially higher for these features (Figure \ref{fig:MI_lat_SED_cond}). These spectral features trace distinct but overlapping stellar evolutionary timescales for the young stellar population \citep[see reviews by][]{kennicutt98, kennicutt12}. On the shortest timescales ($\leq 10$\,Myr), the UV radiation from massive O- and B-type stars ionizes the surrounding gas, driving strong instantaneous emission lines like H~$\upalpha$ (Figure \ref{fig:MI_emission}) \citep{kennicutt98}. On longer timescales ($\sim100$\,Myr), A-type stars dominate the blue optical continuum, producing a prominent Balmer break and strong H~$\updelta$ absorption \citep{bruzual83i, kauffmann03}. For substantially longer timescales, older, cooler stars begin to dominate, driving the metal-line blanketing responsible for the $D_{n}$(4000). The VAE integrates over these different timescales to capture the full unobscured young stellar population. The MI for $z_0$ also peaks strongly across these blue features but drops sharply beyond 5200 \AA\ where the radiation from cooler, older stars dominates. Traversing the latent $z_0$ (Figure \ref{fig:lat_trav}) smoothly maps the physical transition from a star-forming state with strong Balmer emission to a quenched state characterized by absorption features, the $D_{n}$(4000), and a weaker Balmer break \citep{kauffmann03}.

Broadband photometry offers a more limited view because the impact of most emission lines is weak, and the distribution of SF indicator lines across different bands complicates the relationship between SFR and colour \citep{moustakas06, deger25}. While it does capture the 4000 \AA\ region through SFR-sensitive blue colours (traced by the $u-g$ through $z-J$ diagonal in Figure \ref{fig:MI-photometry}), this integration blends distinct physical timescales, giving only a partial picture of the young stellar population.

The VAE, however, does not assign a latent dimension solely to mass-weighted stellar age. In Figure \ref{fig:MI_lat_prop} it is split between $z_0$ and $z_4$, which capture the greatest effects of opposite extremes of the age distribution. The exact spectral signature of the old stellar population is obscured, similarly to how the contribution of these stars is encoded by photometry \citep{papovich01}. The subtle signatures of this population -- an enhanced red continuum and deeper metal absorption lines -- are most heavily impacted by the effects of outshining, and are distributed in the VAE across the latents governing stellar mass, dust, and metallicity (Figure \ref{fig:lat_trav}). 

\subsection{The dust-SFR degeneracy} \label{sec:dust_sfr_deg}
The VAE isolates dust attenuation into latent $z_2$ and SFR into $z_0$ (Figure \ref{fig:MI_lat_prop}). Figure  \ref{fig:MI_lat_SED_cond} and Figure \ref{fig:lat_trav} show that $z_2$ primarily captures the broad 6000$-$7000 \AA\ red slope of the SED, varying the continuum around a ``pinch point'' fixed by $z_4$ (the normalization) without altering other spectral features. By contrast, $z_0$ encodes individual Balmer emission lines. These spectral features separate the two physical processes because they trace distinct mechanisms: the narrow Balmer lines measure instantaneous star formation \citep{kennicutt98}, while the broad optical and NIR slope reflects the continuum reddening caused by dust attenuation \citep{calzetti00} alongside dominant old stellar populations and metals (see Sections \ref{sec:mass} and \ref{sec:dust-metals}). Figure \ref{fig:MI_lat_SED_cond} shows the disentanglement: the MI for $z_0$ and $z_2$ mirror each other. The MI for $z_2$ increases at wavelengths beyond 5000 \AA\ where the red slope dominates, while the MI for $z_0$ decreases.

Broadband photometry, however, struggles to break the degeneracy, particularly at redshifts $z > 0.5$ \citep{kriek13, leja19_uvj, deger25, halder26}. Figure \ref{fig:MI-photometry} shows that optical colours such as $u-g$, $g-r$, and $r-z$ cannot accurately disentangle the dust parameter $\tau_2$ from the SFR, because wide optical filters blend critical features like the 2175~\AA\ UV bump and the 4000~\AA\ break. To break this degeneracy, photometry must rely on the NIR slope. As the red slope shifts out of the $z$-band at $z \simeq 0.5$, NIR bands ($J$, $H$, $K_s$) become essential. These longer-wavelength filters capture the light of cooler, longer-lived stars, strongly attenuated by dust, alongside its own reemission \citep{calzetti00, draine07, salim20}. By capturing these specific effects, the NIR bands allow photometry to isolate the dust component independent of the recent SFR. 

\subsection{The dust-metallicity degeneracy} \label{sec:dust-metals}
The latent dimension $z_2$ governs the broad continuum slope, and so captures not only dust attenuation but also stellar metallicity (Figure \ref{fig:MI_lat_prop}). These two properties create a well-known degeneracy because their impact on the overall slope is so similar: stellar metallicity through line blanketing and dust through continuum attenuation \citep{bell01, salim20}. No latent exclusively encodes stellar metallicity in the rest-frame optical. Its effects are instead captured jointly by the quenched state absorption features of $z_0$, the continuum slope of $z_2$, and the normalization of $z_4$ (Figure \ref{fig:lat_trav}).

Both dust and stellar metallicity also suppress blue flux \citep{worthey94, salim18}, so their resulting red optical/NIR photometric colours ($z-J$, $z-K_s$, $J-K_s$, $K_s-\textit{Ch\,.1}$) are indistinguishable (Figure \ref{fig:MI-photometry}). While photometry can sometimes break this degeneracy by leveraging the rest-frame UV bump to identify young star-forming populations independent of metallicity \citep{salim18}, our optical VAE lacks access to the UV and must rely solely on the shape of the optical continuum.

The VAE separates gas-phase metallicity in $z_1$ and $z_3$ (Figure \ref{fig:MI_lat_prop}) from continuum reddening (dust and stellar metallicity) because these latents encode specific nebular emission lines. Figure \ref{fig:MI_lat_SED_cond} shows that the model carries information about individual metal emission lines such as [O\,\textsc{ii}], [O\,\textsc{iii}], and [N\,\textsc{ii}]. These spectral features, and specifically their ratios to hydrogen lines such as [O\,\textsc{ii}]$/$H~$\upbeta$ and [N\,\textsc{ii}]$/$H~$\upalpha$, are sensitive to the abundance of metals in the interstellar medium (ISM) \citep{pettini04}. The VAE further separates these gas-phase metallicity indicators across $z_1$ and $z_3$ based on their varying ionization potentials. 

By contrast, strong nebular emission lines are blended in photometry, which lacks the spectral resolution required to isolate the line ratios necessary for these gas-phase diagnostics \citep{atek11, maiolino19}. 

\subsection{The two ionization dimensions} \label{sec:IP}
The VAE splits the ISM ionization state into two distinct latent variables $z_1$ and $z_3$ based on the hardness of the radiation. Figure \ref{fig:MI_lat_SED_cond} shows that $z_1$ tracks a soft ionization potential: it captures lines from singly ionized elements such as [O\,\textsc{ii}], [N\,\textsc{ii}], and [S\,\textsc{ii}] with lower ionization potential (Table \ref{tab:emission_lines}). These spectral features trace regions where the radiation field is softer, typical of environments surrounding older, metal-rich stars or lower-mass stars where gas cooling is efficient \citep{kewley02, byler17}. Figure \ref{fig:MI_lat_prop} shows that $Z_{\rm gas}$ and $U_{\rm gas}$ are simultaneously informative about the soft-ionization potential latent $z_1$, because the strength of these singly ionized lines is coupled to both the gas metallicity and the widely varying ionization parameter.

The latent dimension $z_3$ captures the hard ionising radiation field, as it isolates the doubly ionized [O\,\textsc{iii}] and [Ne\,\textsc{iii}] lines (Figure \ref{fig:MI_lat_SED_cond}). Figure \ref{fig:MI_emission} shows that these doubly ionized lines are informative tracers of both gas-phase metallicity and the ionization state. Physically, they trace the high electron temperatures driven by massive stars; strong emission requires a hard radiation field that typically arises when $U_{\rm gas}$ is high and metal-line cooling is inefficient due to low $Z_{\rm gas}$ \citep[see review by][]{kewley19}. Correspondingly, traversing $z_3$ (Figure \ref{fig:lat_trav}) drives variations in the near-UV and blue optical continuum, reflecting the massive stars responsible for this hard radiation. The remainder of the optical spectrum, however, remains modulated by metals and their associated emission lines, save for a broad, smooth uptick in variance at the reddest wavelengths. This red-end variance likely reflects the contribution of the nebular continuum (free-free and free-bound emission), which becomes prominent in highly ionized environments \citep{byler17}. 

By grouping these lines by ionization potential, the VAE bypasses the blending that limits photometry, providing a cleaner diagnostic of the galaxy's physical conditions than colour-colour plots \citep{williams09} or even emission line ratio diagnostics \citep{baldwin81,kauffmann03, kewley06, kewley19} (see Section \ref{sec:discussions}). It also suggests only two degrees of freedom for ionization in the rest-frame optical for typical star-forming galaxies are needed. There is observational evidence for this finding from large-area spectroscopic surveys that suggests that the result is more general. Analysis of optical spectra from SDSS \citep{gyory11} and Dark Energy Spectroscopic Instrument \citep[DESI;][]{khederlarian24} have shown that the information content in emission lines beyond the continuum is low-dimensional. While three or four degrees of freedom are needed for extreme emission line galaxies \citep{kewley02,strom17,strom18,berg21}, our findings in conjunction with these observational results imply that only two are necessary for typical star-forming galaxies.

\begin{table}
\centering
\caption{Key emission lines captured by the VAE and energy (ionization potential; \citealt{emission_lines_NIST}) required to produce them.}
\label{tab:emission_lines}
\begin{tabular}{lcc}
\toprule
line & $\lambda_{\mathrm{em}}$ (\AA) & ionization potential (eV) \\
\midrule
\text{[O\,\textsc{ii}]} 3726    & 3727.1 & 13.62\\
\text{[Ne\,\textsc{iii}]} 3870 & 3869.9 & 40.96\\
H~$\upbeta$ 4861    & 4862.7 & 13.62 \\
\text{[O\,\textsc{iii}]} 5007  & 5008.2 & 35.12\\
\text{[O\,\textsc{i}]} 6302    & 6302.5 & 13.60\\
H~$\upalpha$ 6563   & 6564.6 & 13.60\\
\text{[N\,\textsc{ii}]} 6585   & 6585.3 & 14.53\\
\text{[S\,\textsc{ii}]} 6717   & 6718.3 & 10.36\\
\bottomrule
\end{tabular}
\end{table}

\begin{figure*}
    \centering
	\includegraphics[width=\linewidth]{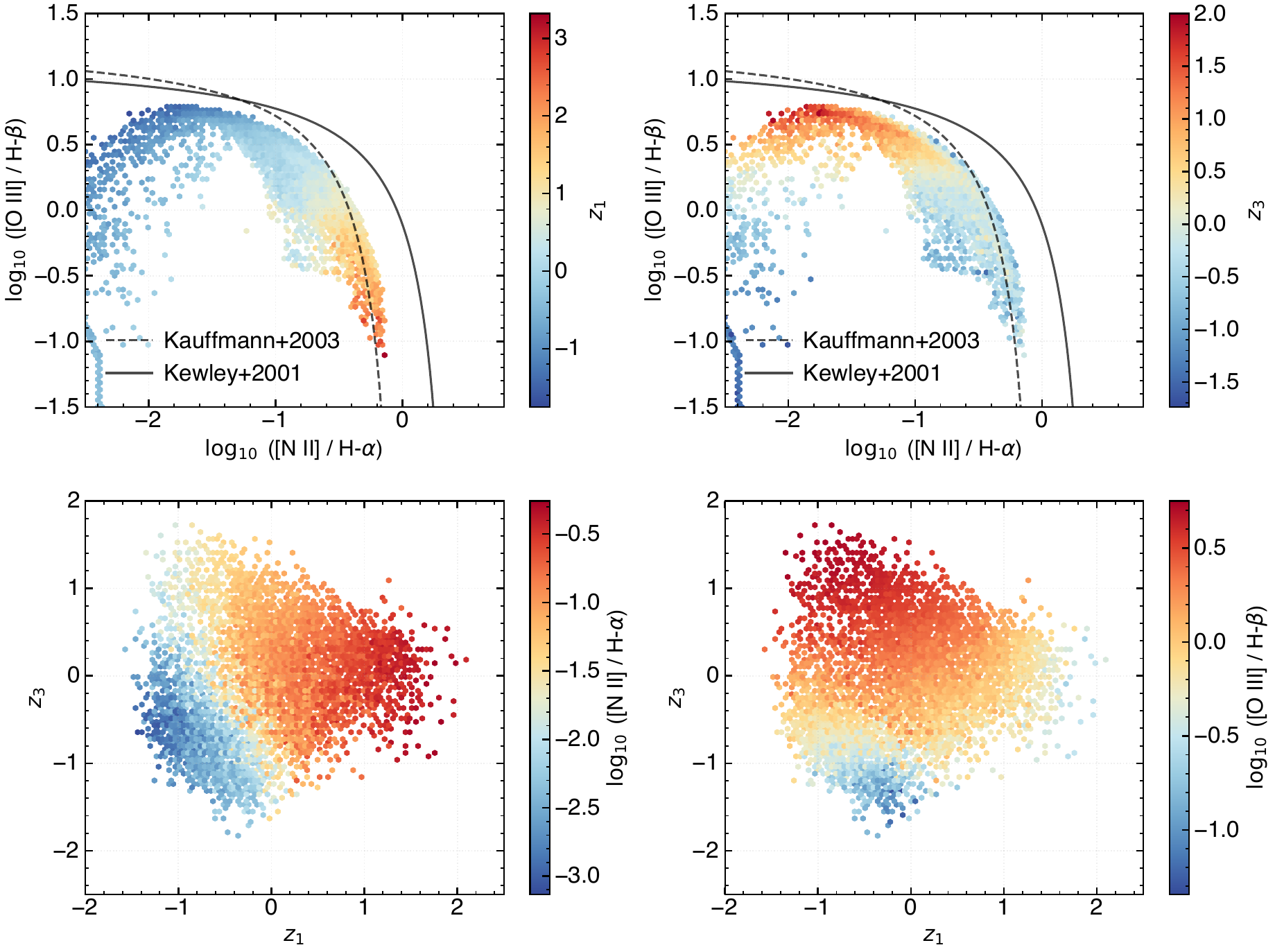}
    \caption{BPT axes against latent $z_1$ and $z_3$. \textit{Top}: BPT diagram coloured by latent $z_1$ (\textit{left}) and latent $z_3$ (\textit{right}). \textit{Bottom}: Latent $z_1$ against latent $z_3$ coloured by the BPT axes $\log_{10}$([N\,\textsc{ii}]$/$H~$\upalpha$) (\textit{left}) and $\log_{10}$([O\,\textsc{iii}]$/$H~$\upbeta$) (\textit{right}). } 
    \label{fig:BPT_lat}
\end{figure*}

\begin{figure*}
    \centering
	\includegraphics[width=\linewidth]{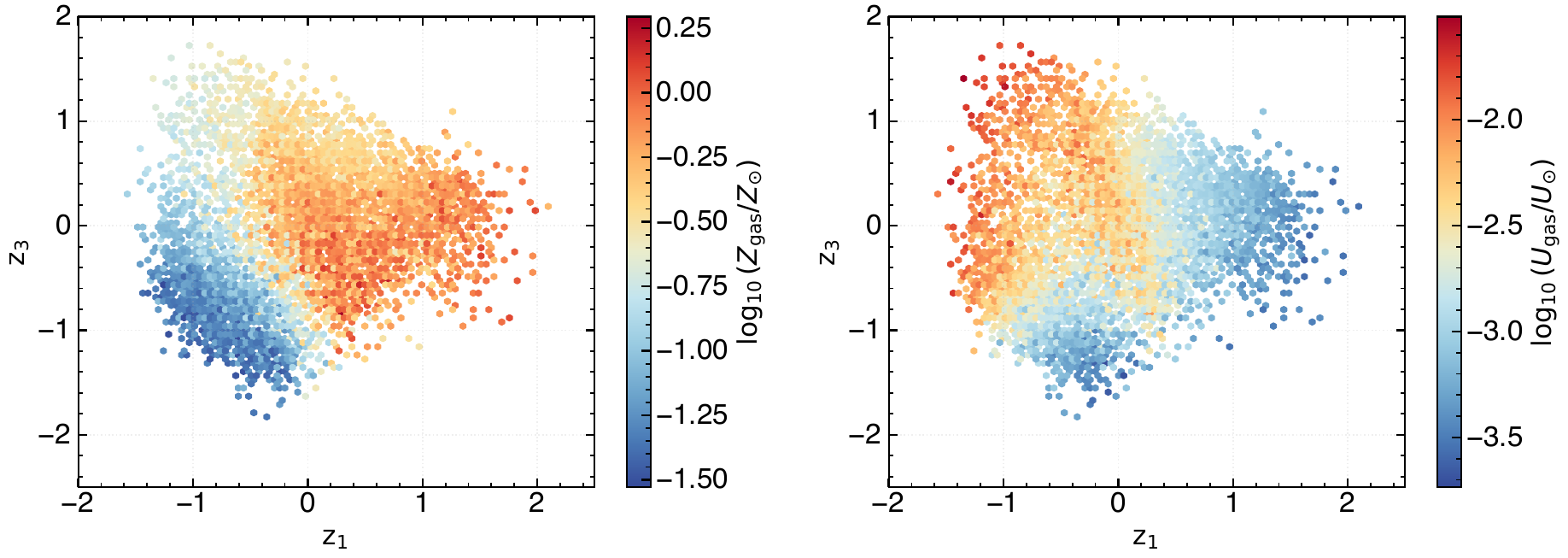}
    \caption{Latent $z_1$ against latent $z_3$ colour coded by gas-phase metallicity (\textit{left}) and gas ionization (\textit{right}).} 
    \label{fig:z1_z3_Zgas_Ugas}
\end{figure*} 

 \section {Discussion} \label{sec:discussions}
The physical conditions of a galaxy's ISM are characterized primarily by its temperature, metallicity and its ionization. Standard optical galaxy evolution diagnostics probe this with ratios of a few nebular emission-lines \citep{osterbrock85,veilleux87,kewley19}, but getting the optimal combination of lines and the number of independent associated physical dimensions needed is a fundamental challenge. Our VAE combines all the information in the SED to give a new diagnostic using the emission-line sensitive latents.

The Baldwin--Phillips--Terlevich (BPT) diagram \citep{baldwin81} is the most widely used diagnostic to distinguish galaxies based on the dominant excitation source (AGN, star-forming galaxies and low-ionization nuclear emission-line regions). We focus on the star forming sequence of the BPT as our SPS model only includes the effect of star formation and not AGN in the optical. The exact position of a galaxy on the BPT star forming sequence is dictated by an interplay of gas-phase metallicity ($Z_{\rm gas}$), ionization parameter ($U_{\rm gas}$), and the hardness of the ionising radiation field (linked to the age). The [N\,\textsc{ii}]$/$H~$\upalpha$ ratio primarily traces $Z_{\rm gas}$ with a secondary inverse dependence on $U_{\rm gas}$ \citep[figure 15]{byler17}. Conversely, the [O\,\textsc{iii}]$/$H~$\upbeta$ ratio traces $U_{\rm gas}$ but is strongly degenerate with $Z_{\rm gas}$ \citep{byler17}. As $Z_{\rm gas}$ drops to subsolar values, reduced metal-line cooling drives up the electron temperature; when coupled with a high $U_{\rm gas}$, this strengthens the [O\,\textsc{iii}] line. However, at extremely low metallicities, the [O\,\textsc{iii}] emission eventually weakens simply due to the diminished absolute oxygen abundance. The diagonal axis of the BPT thus also maps to the age of the galaxies as well as the redshift \citep{kewley13,garg22}.

While the BPT diagram isolates the extremes -- low $Z_{\rm gas}$/high $U_{\rm gas}$ (hard ionising radiation) and high $Z_{\rm gas}$/low $U_{\rm gas}$ (soft ionising radiation) -- it fails to distinguish the physical conditions for galaxies with intermediate values due to strong degeneracies \citep{perez09}. The VAE breaks this degeneracy by combining information from a broader set of features, such as the [S\,\textsc{ii}] line which is sensitive to ionization. 

Figure \ref{fig:BPT_lat} shows the BPT diagram coloured by our VAE latents, $z_1$ and $z_3$. Our galaxies follow the BPT star-forming sequence, and the latents isolate distinct galaxy populations. We have shown previously that these relate to high and low ionization potential and are related to $U_{\rm gas}$ and $Z_{\rm gas}$ (Figure \ref{fig:MI_lat_prop}). By plotting $z_1$ against $z_3$, Figure \ref{fig:z1_z3_Zgas_Ugas} demonstrates that this latent space captures multiple distinct physical regimes: 
\begin{itemize}
    \item $z_1>0.2$: Higher gas-phase metallicity regime. Here $z_1$ is sensitive to $U_{\rm gas}$ while $z_3$ tracks $Z_{\rm gas}$. As $Z_{\rm gas}$ increases, doubly ionized lines become highly sensitive to the metals that dominantly cool the gas cloud's electron temperature. In this high-metallicity regime, $z_3$ spans a narrower range of values, because hard ionising radiation peaks at intermediate gas-phase metallicities before dropping off.
    \item $z_3>-0.8$ and $z_1<0.2$: High ionization regime. Here $z_1$ is sensitive to $Z_{\rm gas}$ and $z_3$ is closely related to the electron temperature as seen by the dependence on [O\,\textsc{iii}] in the bottom right panel of Figure \ref{fig:BPT_lat}.
    \item $z_3<-0.8$ and $z_1<0.2$: Low ionization, low gas-phase metallicity, low mass. Driven by a lack of initial metals and strong outflows that prevent chemical recycling, these galaxies remain largely unenriched and low mass, resulting in weak metal emission lines and low ionization. This population corresponds to the low-mass end of the mass-metallicity relation, where nitrogen is produced predominantly via primary nucleosynthesis rather than the secondary nucleosynthesis pathways that dominate at higher metallicities \citep{considere00}. They appear in the lower left corner of the BPT diagram.
\end{itemize}
The VAE therefore combines information from multiple emission line ratios to provide a diagnostic of the physical regime of the gas cloud.

Our galaxies lie below the \cite{kauffmann03} pure star-formation line, because \texttt{pop-cosmos} does not model the effects of shocks or AGN on optical wavelengths; AGN only enter through the effect of the dusty torus in the NIR \citep{nenkova08i,nenkova08ii,leja18}. The MIR nonetheless carries AGN information that is not degenerate with optical properties (Figure \ref{fig:MI-photometry}). Hence while our optical VAE cannot constrain AGN, a VAE trained on the MIR might.  

\section{Conclusions} \label{sec:Ccl}  
We train a VAE on COSMOS2020-like SEDs drawn from the \texttt{pop-cosmos} galaxy population model, finding that five physical degrees of freedom (latent variables) are necessary and sufficient to describe the rest-frame optical emission. By measuring the MI between the derived latent variables and the underlying SPS parameters, we map the relevant physical processes directly to observable spectral features. This approach isolates the specific intrinsic properties and physical mechanisms required to break well-known observational degeneracies. Recombining these degrees of freedom yields new optical diagnostics for the physical properties of galaxies. 

The \texttt{pop-cosmos} generative model used to create our synthetic data needed 15 SPS parameters (plus redshift) to reproduce the COSMOS2020 UV-NIR photometry~\citep{alsing24,thorp25b}. Our results imply that the rest-frame optical SED in this model can be represented within a significantly lower-dimensional manifold within this high-dimensional parameter space. This framework can also be leveraged to describe galaxies for simulation-based inference (SBI), where the compression provides informative, low-dimensional summary statistics that help keep the inference tractable while retaining interpretability.

Our main conclusions are:
\begin{itemize}
    \item Only five degrees of freedom are required to describe the rest-frame optical SED: stellar mass; young stellar population; dust; and soft and hard ionization potential. Stellar metallicity and galaxy age are not among them; their spectral effects are distributed across the other dimensions rather than independently encoded. Each of the five maps to a clear spectral signature -- the overall SED normalization, the Balmer lines and 4000 \AA\ break region, the broad continuum slope, singly ionized emission lines, and doubly ionized lines, respectively -- features that are often blended together in photometry.

    \item By isolating specific spectral features within distinct latent dimensions, the VAE breaks the observational degeneracies that limit photometric colours. For instance, at intermediate redshifts, optical photometry blends the UV dust bump with the 4000 \AA\ break. By contrast, the VAE separates dust from SFR by independently capturing the broad 6000--7000 \AA\ continuum slope and the narrow Balmer emission-line ratios. Similarly, the VAE disentangles gas-phase metallicity from continuum reddening by isolating specific metallicity-sensitive emission lines, which are otherwise too faint or blended in standard photometric bandpasses. To break these degeneracies observationally, standard photometry requires expanded wavelength coverage -- such as the NIR to decouple dust from SFR, or the rest-frame UV to constrain metallicity. Operating purely in the optical, our VAE bypasses this requirement.
    
    \item By combining the two latent dimensions that capture the soft and hard ionization potentials ($z_1$ and $z_3$), we isolate three distinct physical regimes in which gas-phase metallicity ($Z_{\rm gas}$) and the ionization state ($U_{\rm gas}$) uniquely drive the dominant nebular emission. First, in a high-metallicity regime, efficient metal-line cooling suppresses doubly ionized species, leaving singly ionized lines to primarily trace the ionization parameter. Second, in a subsolar-metallicity regime, reduced cooling allows the ionising radiation to elevate the nebular electron temperature. This drives strong doubly ionized emission, while the strength of singly ionized lines scales with $Z_{\rm gas}$. Finally, we identify a regime characterized by both low metallicity and low ionization, exhibiting overall weak emission features that populate a secondary, low-mass, star-forming sequence on the BPT diagram. That two latent dimensions span this behaviour indicates two independent ionization degrees of freedom in the rest-frame optical for typical star-forming galaxies. That the emission line information of the ISM in typical star-forming galaxies can be represented by a low number of dimensions has been observationally confirmed for SDSS galaxies \citep{gyory11}, and DESI \citep{khederlarian24}.
\end{itemize}

To describe the physical conditions beyond the optical emission, such as for AGN or merger history, wavelengths beyond the rest-frame optical appear necessary. Adding \textit{JWST} NIR and MIR wavebands \citep{jwst_mission} will enable further diagnostics for these additional degrees of freedom. \textit{JWST} and \textit{Euclid} are also pushing the redshift frontier, where distinguishing star-forming from dusty galaxies -- the degeneracy our dust and SFR latents separate -- becomes critical for photo-$z$ estimates \citep{wang24b}. 
Improvements to the SPS modelling framework to account for the additional information from these bands about fainter, higher-redshift and redder objects are likely to reveal additional fundamental degrees of freedom. Wider coverage would also make more of the spectrum accessible to the latents, clarifying the five dimensions identified here. 

\section*{Acknowledgements}
We thank Joy Gong, Joel Leja, and Gary Mamon for useful discussions.

This work has been supported by funding from the European Research Council (ERC) under the European Union's Horizon 2020 research and innovation programmes (grant agreement no.\ 101018897 CosmicExplorer). This work has been enabled by support from the research project grant ‘Understanding the Dynamic Universe’ funded by the Knut and Alice Wallenberg Foundation under Dnr KAW 2018.0067. HVP was additionally supported by the G\"{o}ran Gustafsson Foundation for Research in Natural Sciences and Medicine.

This work was performed using resources provided by the Cambridge Service for Data Driven Discovery (CSD3) operated by the University of Cambridge Research Computing Service (\url{www.csd3.cam.ac.uk}), provided by Dell EMC and Intel using Tier-2 funding from the Engineering and Physical Sciences Research Council (capital grant EP/T022159/1), and DiRAC funding from the Science and Technology Facilities Council (\url{www.dirac.ac.uk}).

\textit{Software:} \texttt{NumPy} \citep{harris20}; \texttt{Pandas} \citep{mckinney10}; \texttt{SciPy} \citep{virtanen20}; \texttt{Matplotlib} \citep{hunter07}; \texttt{astropy} \citep{astropy13,astropy18,astropy22}; \texttt{PyTorch} \citep{paszke19}; \texttt{Speculator} \citep{alsing20}; \texttt{torchdiffeq} \citep{chen18}; \texttt{Prospector} \citep{johnson21}; \texttt{FSPS} v3.2 \citep{conroy09, conroy10a, conroy10b}; \texttt{python-fsps} v0.4.1 \citep{pythonfsps}; \texttt{sedpy} \citep{sedpy}; \texttt{tqdm} \citep{tqdm}; \texttt{GMM-MI} \citep{piras23}. 

\section*{Data Availability}
The galaxy SPS parameters drawn from \texttt{pop-cosmos}, and their associated forward modelled observed frame photometry can be found on Zenodo and are described in \citet{thorp25b} and \citet{deger25}. We use the SPS parameters and model photometry from v1.2.0 of \verb!mock_catalog_Ch1_26.h5!, and provide an updated v1.4.1 of the Zenodo record \citep{thorp26_mock} that includes rest-frame SEDs for the full \verb!mock_catalog_Ch1_26.h5! in a new file: \verb!mock_rest_seds_Ch1_26.h5!.

\section*{Author Contributions}
 We outline the different contributions below using keywords based on the Contribution Roles Taxonomy (CRediT; \citealp{brand15}).
 \textbf{BVdB:} conceptualization; 
formal analysis; methodology; 
software; 
visualization; 
validation; 
investigation; 
data curation; 
writing -- original draft, review \& editing.
\textbf{SD:} conceptualization; investigation; methodology; validation; software; supervision; writing -- review \& editing.
\textbf{HVP:} conceptualization; investigation; methodology; validation; visualization; funding acquisition; project administration; supervision; writing -- review \& editing.
\textbf{ST:} software; data curation; writing -- review \& editing.
\textbf{DJM:} 
visualization;
writing -- review \& editing.
\textbf{AH:} writing -- review.
\textbf{BL:} 
writing -- review \& editing.
\textbf{MNT:} 
writing -- review \& editing.
\textbf{GJ:} 
software.

\textit{Declaration of LLM use:} Anthropic’s Claude Opus 4.8 and Google Gemini 3.1 Pro have been used to assess flow and redundancy of text and proofreading as well as suggestions for condensing. GitHub Copilot Pro has also been used to correct syntactical errors, suggest code optimizations, and help produce plots.



\bibliographystyle{mnras}
\bibliography{pop-cosmos}



\appendix
\section{Training strategy} \label{sec:annealing}
\begin{figure}
	\includegraphics[width=\linewidth]{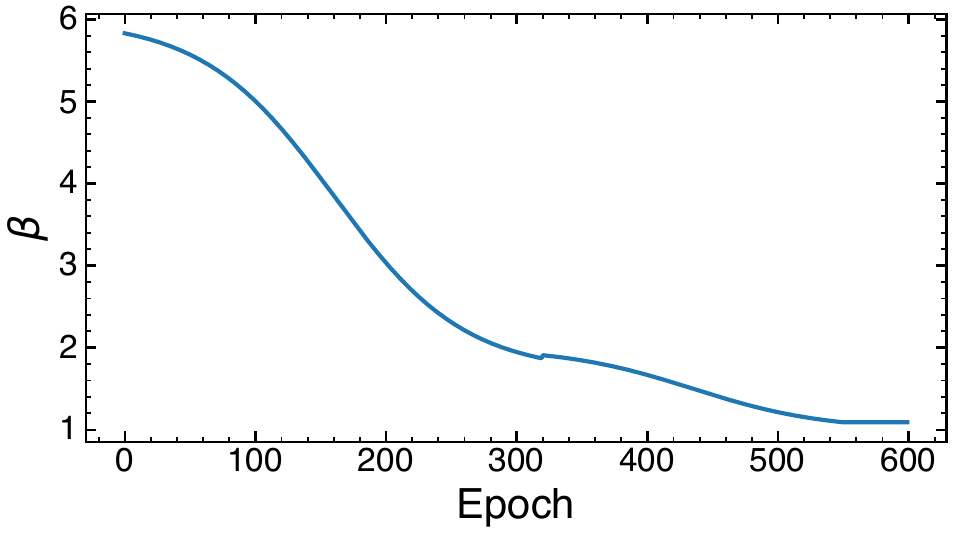}
    \caption{The $\beta$-annealing schedule during the training of the VAE.}
    \label{fig:beta_annealing}
\end{figure}

We apply a three-stage $\beta$-annealing process to train our model, aiming to reproduce the baseline 2 per cent accuracy of the $\beta=0$ trained VAE. We modify the value of $\beta$ over the training epochs according to the following sigmoid-like function:
\begin{equation} \label{eq:annealing}
    \beta'(t) = \beta_i + \frac{\beta_f -\beta_i}{1+\exp[-r(t-t_c)]},
\end{equation}
where $t$ is the current epoch, $t_c$ is the epoch of maximal annealing, and $r$ controls the rate of change. The slope of the annealing curve is decreased in each subsequent stage, as visualized in Figure \ref{fig:beta_annealing}. $\beta$ is decreased from 6 to 1.9 over the first 320 epochs with $t_c=60$ and $r=0.02$ in the first stage. During the second stage $\beta$ goes from 1.9 to 1.1 for 220 epochs with the same $t_c$ and $r$. Finally in the third stage $\beta$ is kept constant for 60 epochs. 
By gradually decreasing $\beta$, the KL divergence term becomes increasingly less important in the loss function, encouraging the model to capture all the information in the SEDs, while the initial stages prioritize the independence of the latent variables. The rate $r$ and $t_c$ are chosen such that the slope decreases gradually enough to reach the desired value of $\beta$ after the given number of epochs. 

The other hyperparameters of our VAE are tuned to ensure high-fidelity reconstructions of the SEDs while mitigating overfitting. The VAE is trained for 600 epochs with the $\beta$-annealing schedule described above with a batch size of 64. We optimize the network using the \texttt{Adam} optimizer \citep{kingma14}, setting the initial learning rate to $10^{-4}$ and weight decay to $10^{-5}$. To stabilize convergence, we apply a step learning rate scheduler that decays the learning rate by a factor of 0.5 every 20 epochs.  


\bsp	
\label{lastpage}
\end{document}